\documentclass[a4paper]{article}
\usepackage{graphicx}
\usepackage{onecolceurws}
\usepackage[cp1251]{inputenc}
\usepackage{amssymb,amsmath}

\newcommand\bovermat[2]{%
  \makebox[0pt][l]{$\smash{\overbrace{\phantom{%
    \begin{matrix}#2\end{matrix}}}^{\text{#1}}}$}#2}

\title{Two-Dimensional Control and Assurance
of Data Integrity in Information Systems
Based on Residue Number System Codes
and Cryptographic Hash Functions}

\author{
Sergey Dichenko \\ University Teachers \\ Institute of Computer Systems and \\ Information Security of Kuban \\ State Technological University \\
                Krasnodar, Russia \\ dichenko.sa@yandex.ru
\and
Oleg Finko \\ Professor \\ Institute of Computer Systems and \\ Information Security of Kuban \\ State Technological University \\
                Krasnodar, Russia \\ ofinko@yandex.ru
}

\institution{}

\begin{document}
\maketitle

\begin{abstract}
The method of two-dimensional control and assurance of data integrity with the
possibility of their recovery for information systems operating under
conditions of random errors as well as errors generated through
deliberate actions of the attacker is proposed. The data recovery procedure is
based on the application of the mathematical apparatus of redundant
residue number system codes (RNSC), and the control (verification of the recovered
data validity (reliability, accuracy)) of data integrity is performed by
means of cryptographic methods.
\end{abstract}
\vskip 32pt

\section{Introduction}

At present, users of various information systems are facing the tasks of protecting the data processed in them.
One of the measures to ensure the security of data processed in information systems is the protection of their
integrity \cite{ISO-17799}.

The problem solution of data integrity protection becomes especially urgent during the operation of widely created
data processing centers when using different processing facilities in their composition with different building
structures and operating principles under conditions of both random errors and errors generated through deliberate
actions of an attacker (unauthorized data modification (for example, through the action of malicious code)
or the failure of a part of the media (for example, individual cells, sectors)).

The challenge of data integrity protection is complicated because of its complexity, as it involves not only data
integrity control, but also its provision, which means the restoration of data whose integrity has been violated
for some reason.

There are various ways of solving the problem of control and assurance of data integrity, among which the following
are of the greatest interest.

\section{Analysis of existing solutions for control and assurance of data integrity}

There are known ways to control the data integrity by calculating checksum values and comparing them with reference
values, as well as methods based on the use of cryptographic methods: key and keyless hashing, means of electronic
signature \cite{Knuth-vol3, Menezes-Handbook, Biham-hash, Bellare-NMAC}. The disadvantage
of these methods is the lack of the ability to insure their integrity without introducing an additional data
recovery mechanism.

There are known ways to ensure the integrity of data through the use of various types of reservation
(using hardware or software implementation of RAID technology (Redundant Array of Independent Disks)
(RAID arrays)), duplication methods, redundant coding methods \cite{Henry-Hacker, Morelos-Zaragoza-Art, Hamming-Coding}.
The disadvantage of these methods is high redundancy.

For this article the two-dimensional methods of excess coding in the residue number system \cite{Vasyl-Yatskiv} taken by us as a prototype will be of particular importance many-dimensional, in particular.

The presented solutions show that some of the methods allow to control the data integrity by comparing
the values of the reference and calculated hash-codes of the hash function (checksums) when requesting
the use of processed or stored data, but the lack of a mechanism for their recovery does not
allow their integrity assurance. Other methods, on the other hand, provide data integrity by restoring them,
for example, from a backup copy, but their practical use without data integrity control is ineffective.
Individual methods allow for control and ensure the data integrity, however, of valuable high
redundancy.

The most popular solutions are the complex protection of data integrity associated with the
simultaneous solution of control tasks and ensuring data integrity, which is achieved by
consistently applying first the cryptographic transformation to data, and then applying
the technology of data backup.

At the same time, data integrity protection is relevant both for systems of RAID type,
where all media are located in one constructive block, and for distributed storage systems,
that is, for network storage.

Thus, in order to protect the data integrity, when considering this notion in a complex,
it is necessary to aggregate existing solutions. Combining the known methods in one allows
you to control and ensure the data integrity.

\section{Choosing ways to control the data integrity and recovery to share them while
ensuring integrity}

A method \cite{Application-0081048} is known where, before writing to a RAID array (after reading), the data is encrypted
(decrypted) by a dedicated device connected to the PCI-tyre, the encryption key being read from an
external storage device and/or requested from the user. In \cite{Patent-8209551}, before writing to the array,
the data is divided into several segments, after which the checksums are calculated separately
from the data from each segment. The data segments and checksums are further distributed over
the disks of the RAID array.

In \cite{Patent-7752676}, a method of protecting data in a network storage is proposed, where a user's request
for reading (writing) data first passes the authorization procedure, and only if the operation
is allowed, the data on the network storage is decrypted (encrypted) accordingly. The keys
of encryption (decryption) are stored on the client side.

Another version of combined protection is proposed in \cite{Application-0107103}, where the data is stored in the cloud,
and the encryption module is stored not on the client side, but on the side of the cloud storage
provider. This solution is intended, as a rule, to protect the backup copies of data in the cloud,
although the original data is stored on the client side in its original form. In order to protect
the data, the data file is first divided into parts, and then each part is transformed using
a cryptographic algorithm and written to one or more media in the cloud. Protection is provided
when data is lost on the client side. In this case, the backup is restored from the cloud.

The disadvantage of the presented combined methods is the high redundancy, as well as the lack
of the possibility, without the introduction of an additional monitoring mechanism, to verify
the validity (reliability, accuracy) of the recovered data while ensuring their integrity.

In order to eliminate the drawbacks of the known combined methods, a solution is proposed in which
cryptographic methods are chosen to perform data integrity control, in particular, a hash function
designed specifically for this purpose, and the data recovery procedure is performed by using
redundant residue number system codes, the application of the mathematical apparatus of which
allows to provide minimal redundancy, and most importantly, provides, when used together with
cryptographic methods the construction of unique scheme which allows to verify the validity
(reliability, accuracy) of recovered data while ensuring its integrity in case of violation.

\section{Structural-parametric synthesis of the system of parallel control and assurance of data integrity}

For control and integrity purposes, the data blocks $M_{i}$ ($i=1,\,2,\,\ldots,\,n$), to
be protected are represented in the form of sub-blocks of fixed length
$M_{i}=\{m_{i,\,1} || m_{i,\,2} || \ldots || m_{i,\,n}\}$, where $||$~--- is the concatenation operation,
$n$~--- is the number of data blocks $M_{i}$, to be protected, and also fixed-length sub-blocks in
each data block under consideration $M_{i}$. And the length of the data blocks~$M_{i}$ equals 512~bits.

Obtaining the matrix $\mathbf{W}$:

\begin{align*}
\mathbf{W} = \begin{matrix}
\begin{bmatrix}
~\bovermat{$~~~~512 \ \text{bits}~~~~$}
{m_{1,\,1} & m_{1,\,2} & \cdots & m_{1,\,n}} \ \ \\[0.2em]
m_{2,\,1}  & m_{2,\,2} & \cdots & m_{2,\,n}  \ \ \\[0.2em]
\vdots   & \vdots  & \ddots & \vdots   \ \ \\[0.2em]
m_{n,\,1}  & m_{n,\,2} & \cdots & m_{n,\,n}  \ \ \\[0.2em]
\end{bmatrix}
\begin{aligned}
&\left.\begin{matrix}
    \\[0.5em]
    \\[0.5em]
    \\[0.5em]
\end{matrix} \right\} %
512 \ \text{bits}\\
\end{aligned}.
\end{matrix}
\end{align*}

To implement integrity control, a hash function is applied to the data blocks~$M_{i}$, the construction rules of which are defined in \cite{ISO-14888--1}.
%
%
The received hash-codes $S_{i}$ hash functions $h(M_{i})$ from data blocks $M_{i}$ will be the reference codes,
we obtain the matrix $\mathbf{\Psi}$:

\begin{align*}
\mathbf{\Psi} = \begin{matrix}
\begin{bmatrix}
~\bovermat{$~~~~512 \ \text{bits}~~~~$}
{m_{1,\,1} & m_{1,\,2} & \cdots & m_{1,\,n}} & \rightarrow & \bovermat{$512 \ \text{bits}$}{s_{1,\,1} & s_{1,\,2} & \cdots & s_{1,\,n}} \ \ \\[0.2em]
m_{2,\,1}  & m_{2,\,2} & \cdots & m_{2,\,n}  & \rightarrow & s_{2,\,1} & s_{2,\,2} &\cdots & s_{2,\,n} \ \ \\[0.2em]
\vdots   & \vdots  & \ddots & \vdots   & \cdots & \vdots & \vdots & \ddots & \vdots  \ \ \\[0.2em]
m_{n,\,1}  & m_{n,\,2} & \cdots & m_{n,\,n}  & \rightarrow & s_{n,\,1} & s_{n,\,2} & \cdots & s_{n,\,n} \ \ \\[0.2em]
\end{bmatrix}
\begin{aligned}
&\left.\begin{matrix}
    \\[0.5em]
    \\[0.5em]
    \\[0.5em]
\end{matrix} \right\} %
512 \ \text{bits}\\
\end{aligned},
\end{matrix}
\end{align*}
where $S_{i}=h(M_{i})$; $S_{i}=\{s_{i,\,1} || s_{i,\,2} || \ldots || s_{i,\,n}\}$.

Now consider the data blocks $M_{j}$ ($j=1,\,2,\,\ldots,\,n$) represented by sub-blocks
$m_{1,\,1},\,m_{2,\,1},\,\ldots,\,m_{n,\,1}$; $m_{1,\,2},\,m_{2,\,2},\,\ldots,\,m_{n,\,2}; \ldots;
m_{1,\,n},\,m_{2,\,n},\,\ldots,\,m_{n,\,n}$. The sub-blocks of the $m_{i,\,j}$ data blocks under consideration $M_{j}$
are interpreted as the minimum nonnegative deductions from the generically ordered, mutually simple modules $p_{i,\,j}$,
and form an information super-block of the RNSC.

As a result of the base extension, we obtain redundant sub-blocks $m_{n+1,\,1}$,
$m_{n+2,\,1},\,\ldots,\, m_{k,\,1}$; $m_{n+1,\, 2},\, m_{n+2,\, 2},\, \ldots,\, m_{k,\,2};\, \ldots ;\, m_{n+1,\,n},\, m_{n+2,\, n},\, \ldots,\, m_{k,\, n}$, the set of which together with the sub-blocks forming
a single super-block of elements form a code vector of the RNSC.

We get the matrix $\mathbf{\Upsilon}$ with redundant sub-blocks of the code vector of the RNSC:

\begin{align*}
\mathbf{\Upsilon} = \begin{matrix}
\begin{bmatrix}
~\bovermat{$~~~~512 \ \text{bits}~~~~$}
{m_{1,\,1}   & m_{1,\,2}   &\cdots & m_{1,\,n}}   & \bovermat{$512 \ \text{bits}$}{s_{1,\,1} & s_{1,\,2} &\cdots & s_{1,\,n}} \\[0.2em]
m_{2,\,1}    & m_{2,\,2}    &\cdots & m_{2,\,n}    & s_{2,\,1} & s_{2,\,2} & \cdots & s_{2,\,n} \\
\vdots     & \vdots     &\ddots & \vdots     & \vdots  & \vdots  & \ddots & \vdots  \\
m_{n,\,1}    & m_{n,\,2}    &\cdots & m_{n,\,n}    & s_{n,\,1} & s_{n,\,2} & \cdots & s_{n,\,n} \\
\downarrow  & \downarrow  & \vdots & \downarrow  & \ & \ & \ & \ \\
m_{n+1,\,1}  & m_{n+1,\,2}  & \cdots & m_{n+1,\,n}  & \ & \ & \ & \ \\
m_{n+2,\,1}  & m_{n+2,\,2}  & \cdots & m_{n+2,\,n}  & \ & \ & \ & \ \\
\vdots     & \vdots     & \ddots & \vdots     & \ & \ & \ & \ \\
m_{k,\,1}   & m_{k,\,2}   & \cdots & m_{k,\,n}   & \ & \ & \ & \ \\
\end{bmatrix}
\begin{aligned}
  &\left.\begin{matrix}
    \\[0.5em]
    \\[0.5em]
    \\[0.5em]
  \end{matrix} \right\} %
  512 \ \text{bits}.\\
  \\
  \\
  \\
  \\
  \\
  \\
    \end{aligned}
\end{matrix}
\end{align*}

We add the $i$-th sub-blocks of hash-codes $S_{i}$ with $j$-th redundant sub-blocks of data blocks $M_{j}^{*}$
of the code vector of the residue number system codes:
\begin{align*}
  G_{i}=S_{i}\oplus M_{j}^{*}=(s_{i,\,1}\oplus m_{n+1,\,j}; s_{i,\,2}\oplus m_{n+2,\,j}; \ldots; s_{i,\,n}\oplus m_{k,\,j}),
\end{align*}
where the sign ``$\oplus$'' denotes the summation in the Galois field GF($2$), $i=j$,
$S_{i}=[\begin{array}{cccc}
            s_{i,\,1} & s_{i,\,2} & \ldots & s_{i,\,n}
            \end{array}]$, $M^{*}_{j}=[\begin{array}{cccc}
            m_{n+1,\,j} & m_{n+2,\,j} & \ldots & m_{k,\,j}
            \end{array}]^{\top}$, $G_{i}=[\begin{array}{cccc}
            g_{i,\,1} & g_{i,\,2} & \ldots & g_{i,\,n}
            \end{array}]$.

We obtain the matrix $\mathbf{\Omega}$:

\begin{align}\label{1}
\mathbf{\Omega} = \begin{matrix}
\begin{bmatrix}
~\bovermat{$~~~~512 \ \text{bits}~~~~$}
{m_{1,\,1} & m_{1,\,2} & \cdots & m_{1,\,n}} & \bovermat{$512 \ \text{bits}$}{g_{1,\,1} & g_{1,\,2} & \cdots & g_{1,\,n}} \ \ \\[0.2em]
m_{2,\,1}  & m_{2,\,2} & \cdots & m_{2,\,n}  & g_{2,\,1} & g_{2,\,2} &\cdots & g_{2,\,n} \ \ \\[0.2em]
\vdots   & \vdots  & \ddots & \cdots & \vdots & \vdots & \ddots & \vdots  \ \ \\[0.2em]
m_{n,\,1}  & m_{n,\,2} & \cdots & m_{n,\,n}  & g_{n,\,1} & g_{n,\,2} & \cdots & g_{n,\,n} \ \ \\[0.2em]
\end{bmatrix}
\begin{aligned}
&\left.\begin{matrix}
    \\[0.5em]
    \\[0.5em]
    \\[0.5em]
\end{matrix} \right\} %
512 \ \text{bits}\\
\end{aligned}.
\end{matrix}
\end{align}

\begin{figure}[ht]
\begin{center}
\includegraphics[height=7.0cm]{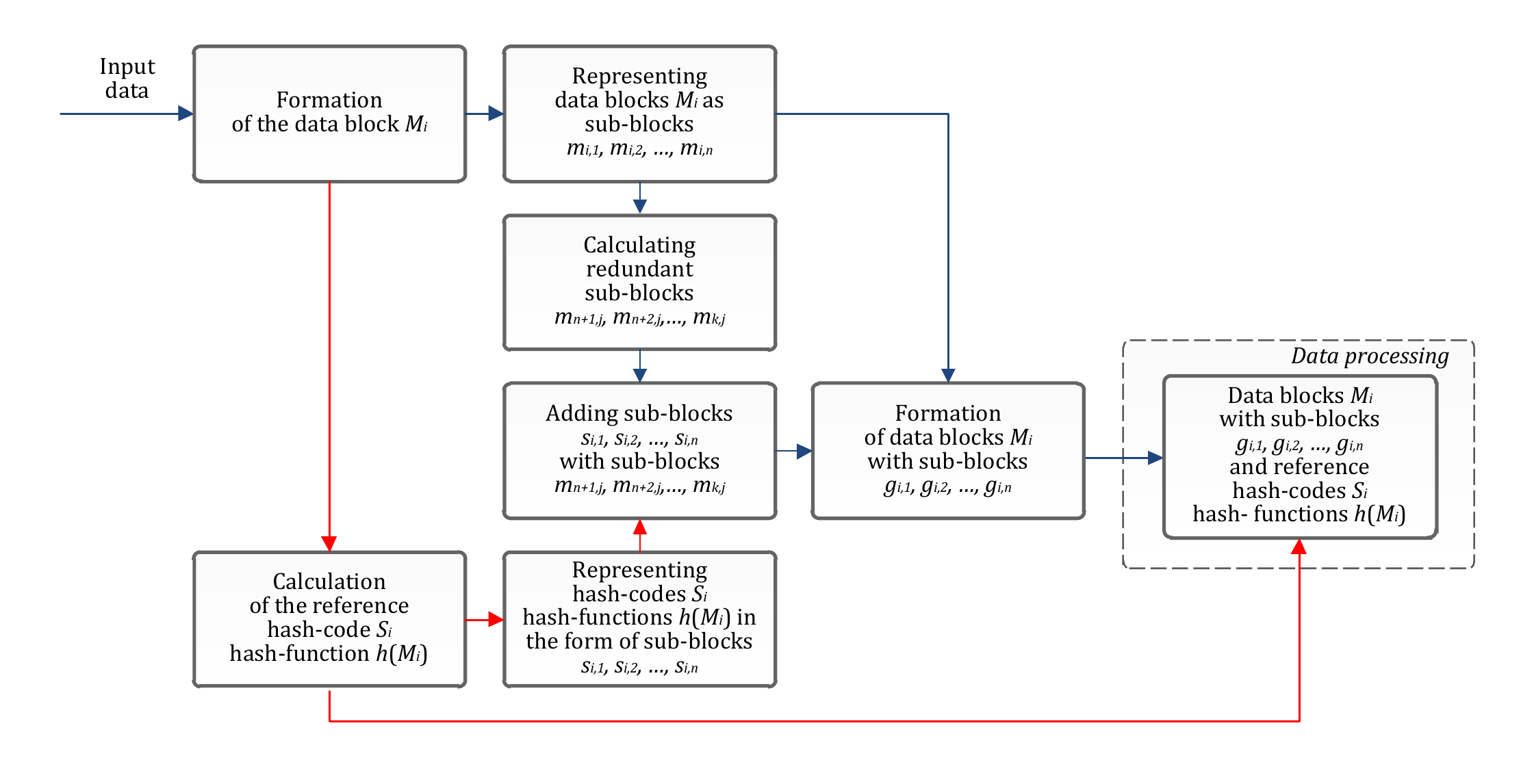}
\caption{A diagram explaining the preparatory stage of the system construction}
\label{fig_1}
\end{center}
\end{figure}

At the end of the preparatory stage of the construction of the system (Figure \ref{fig_1}), the data subject to protection
is presented in the form (\ref{1}), which will allow control and ensuring their integrity.

\section{Procedure for control of data integrity}

When requesting the use of data (the main stage) to be protected, they are control for their integrity,
which can be ensured by performing the base extension the information super-block of the residue number system codes \cite{Bajard-RNS, Bajard-Efficient RNS},
with this redundant sub-blocks are created $m^{\prime}_{n+1,\,1},\,m^{\prime}_{n+2,\,1},\,\ldots,\,m^{\prime}_{k,\,1}; m^{\prime}_{n+1,\,2},\,m^{\prime}_{n+2,\,2},\,\ldots,
m^{\prime}_{k,\,2}; \ldots$; $m^{\prime}_{n+1,\,n},\,m^{\prime}_{n+2,\,n},\,\ldots,\,m^{\prime}_{k,\,n}$ of the data blocks $M_{j}^{*\prime}$
of the code vector of the residue number system codes, where ``$\bullet^{\prime}$'' denotes that changes could occur in sub-blocks
$m^{\prime}_{1,\,1},\,m^{\prime}_{1,\,2},\,\ldots,\,m^{\prime}_{1,\,n}; m^{\prime}_{2,\,1},\,m^{\prime}_{2,\,2},\,\ldots,\,m^{\prime}_{2,\,n}; \ldots$;
$m^{\prime}_{n,\,1},\,m^{\prime}_{n,\,2},\,\ldots,\,m^{\prime}_{n,\,n}$ of~data blocks $M^{\prime}_{i}$.

The matrix $\mathbf{\Omega}$ with the redundant sub-blocks of the code vector of the RNSC takes the form:

\begin{align*}
\mathbf{\Omega^{\prime}} = \begin{matrix}
\begin{bmatrix}
~\bovermat{$~~~~512 \ \text{bits}~~~~$}
{m^{\prime}_{1,\,1}   & m^{\prime}_{1,\,2}   &\cdots & m^{\prime}_{1,\,n}}   & \bovermat{$512 \ \text{bits}$}{g_{1,\,1} & g_{1,\,2} &\cdots & g_{1,\,n}} \\[0.2em]
m^{\prime}_{2,\,1}    & m^{\prime}_{2,\,2}    &\cdots & m^{\prime}_{2,\,n}    & g_{2,\,1} & g_{2,\,2} & \cdots & g_{2,\,n} \\
\vdots     & \vdots     &\ddots & \vdots     & \vdots  & \vdots  & \ddots & \vdots  \\
m^{\prime}_{n,\,1}    & m^{\prime}_{n,\,2}    &\cdots & m^{\prime}_{n,\,n}    & g_{n,\,1} & g_{n,\,2} & \cdots & g_{n,\,n} \\
\downarrow  & \downarrow  & \vdots & \downarrow  & \ & \ & \ & \ \\
m^{\prime}_{n+1,\,1}  & m^{\prime}_{n+1,\,2}  & \cdots & m^{\prime}_{n+1,\,n}  & \ & \ & \ & \ \\
m^{\prime}_{n+2,\,1}  & m^{\prime}_{n+2,\,2}  & \cdots & m^{\prime}_{n+2,\,n}  & \ & \ & \ & \ \\
\vdots     & \vdots     & \ddots & \vdots     & \ & \ & \ & \ \\
m^{\prime}_{k,\,1}   & m^{\prime}_{k,\,2}   & \cdots & m^{\prime}_{k,\,n}   & \ & \ & \ & \ \\
\end{bmatrix}
\begin{aligned}
  &\left.\begin{matrix}
    \\[0.5em]
    \\[0.5em]
    \\[0.5em]
  \end{matrix} \right\} %
  512 \ \text{bits}.\\
  \\
  \\
  \\
  \\
  \\
  \\
    \end{aligned}
\end{matrix}
\end{align*}

We perform the inverse transformation:
\begin{align*}
  S^{\prime}_{i}=G_{i}\oplus M_{j}^{*\prime}=(g_{i,\,1}\oplus m^{\prime}_{n+1,\,j}; g_{i,\,2}\oplus m^{\prime}_{n+2,\,j}; \ldots; g_{i,\,n}\oplus m^{\prime}_{k,\,j}),
\end{align*}
where $S^{\prime}_{i}=[\begin{array}{cccc}
            s^{\prime}_{i,\,1} & s^{\prime}_{i,\,2} & \ldots & s^{\prime}_{i,\,n}
            \end{array}]$, $M^{*\prime}_{j}=[\begin{array}{cccc}
            m^{\prime}_{n+1,\,j} & m^{\prime}_{n+2,\,j} & \ldots & m^{\prime}_{k,\,j}
            \end{array}]$, $i=j$.

Compare the values of the hash-codes obtained~$S^{\prime}_{i}$ hash function~$h(M^{\prime}_{i})$ to the values
of the previously calculated reference hash-codes~$S_{i}$ hash function~$h(M_{i})$.
Based on the results of the comparison, let's make a conclusion:
\begin{itemize}
      \item [$\blacktriangleright $] about the absence of violation of data integrity, at $S^{\prime}_{i}=S_{i}$;
      \item [$\blacktriangleright $] about data integrity violation, when $S^{\prime}_{i}\neq S_{i}$.
\end{itemize}

\section{Procedure of ensuring the data integrity}

If the values of the hash-codes of the hash function compared with each other are different,
which will be characterized by the occurrence of an error (violation of integrity) in the data
being processed, we shall perform its localization.

The localization of the detected error (sub-blocks $\tilde{m}_{i,\,j}$ with integrity violation)
is performed initially on the rows of the matrix $\mathbf{\Omega^{\prime}}$ (the $i$-th data block with the
integrity violation, which includes the sub-block $\tilde{m}_{i,\,j}$ is determined), and then on the columns
(the $j$-th data block with integrity violation, which includes the sub-block $\tilde{m}_{i,\,j}$ is determined).

A data block $\tilde{M}_{i}$ with integrity violation, whose sub-blocks are located along the row of the matrix
$\mathbf{\Omega^{\prime}}$, is determined from the results of a comparison of the calculated and reference hash-codes
of the hash function. A data block $\tilde{M}_{j}$ with an integrity violation whose sub-blocks are arranged along
the column of the matrix $\mathbf{\Omega^{\prime}}$ is determined by means of a mathematical apparatus of redundant
RNSC based on the fundamental provisions of the Chinese remainder theorem.

In accordance with the mathematical apparatus of Residue Number System (RNS) \cite{Bajard-RNS}, in which the tested data block~$M_{j}$
will be interpreted as a nonnegative integer~$A_{j}$ unambiguously represented by a set of residues on RNS
basis $p_{1,\,j},\,p_{2,\,j},\,\ldots,\,p_{n,\,j}<p_{n+1,\,j}<\ldots<p_{k,\,j}$:
\begin{align*}\
   A_{j}=(\alpha_{1,\,j},\,\alpha_{2,\,j},\,\ldots,\,\alpha_{n,\,j},\,\alpha_{n+1,\,j},\,\ldots,\,\alpha_{k,\,j}),
\end{align*}
where $P_{n,\,j}=p_{1,\,j}p_{2,\,j}\ldots p_{n,\,j}>A_{j}$; $\alpha_{i,\,j}=|A|_{p_{i,\,j}}$; $|\bullet|_{p}$~---
is the smallest nonnegative residue of the number~``$\bullet$'' modulo $p$; $j=1,\,2,\,\ldots,\,n,\,n+1,\,\ldots,\,k$; $i=1,\,2,\,\ldots,\,n$;
$p_{1,\,j},\,p_{2,\,j},\,\ldots,\,p_{n,\,j}<p_{n+1,\,j}<\ldots<p_{k,\,j}$~--- are pairwise simple.

The resulting residues $\alpha_{i,\,j}$ will be interpreted as sub-blocks $m_{i,\,j}$ of the data block $M_{j}$,
that is, the remnants of the RNS $\alpha_{1,\,j},\,\alpha_{2,\,j},\,\ldots,\,\alpha_{n,\,j}$ will be interpreted as sub-blocks
$m_{1,\,j},\,m_{2,\,j},\,\ldots,\,m_{n,\,j}$ and will be considered informational (informational group $n$ sub-blocks), and
$\alpha_{n+1,\,j},\,\ldots,\,\alpha_{k,\,j}$~--- interpreted as sub-blocks $m_{n+1,\,j},\,\ldots,\,m_{k,\,j}$ and considered as control (redundant)
(control (redundant) group $(k-n)$ sub-blocks). The RNS itself is in this case extended, where $P_{k,\,j}=P_{n,\,j}p_{n+1,\,j}\ldots p_{k,\,j}$,
and covers the complete set of states represented by all $k$ deductions. This area will be the full range of the RNS
$[0,\,P_{k,\,j})$ and consist of a working range $[0,\,P_{n,\,j})$, where $P_{n,\,j}=p_{1,\,j}p_{2,\,j}\ldots p_{n,\,j}$, is defined by nonredundant of the RNS bases
(sub-blocks $m_{1,\,j},\,m_{2,\,j},\,\ldots,\,m_{n,\,j}$), and a range $[P_{n,\,j},\,P_{k,\,j})$ defined by redundant of the RNS bases
(sub-blocks $m_{n+1,\,j},\,\ldots,\,m_{k,\,j}$) and representing invalid area. This means that operations on the number $A_{j}$
are performed in the range $[0,\,P_{k,\,j})$, and if the result of the RNS operation goes beyond the $P_{n,\,j}$,
then there is a conclusion about the calculation error. Checking this rule allows you to localize the error in the data block
$\tilde{M}_{j}$ of the matrix $\mathbf{\Omega^{\prime}}$.

\textbf{Example 1}

Choose a base system $p_{1}=2,\, p_{2}=3,\, p_{3}=5,\, p_{4}=7$ for which the operating range is
$P_{4}=p_{1}p_{2}p_{3}p_{4}=\\
= 2\cdot3\cdot5\cdot7=210$. Then introduce the control bases $p_{5}=11,\, p_{6}=13$,
then the full range is defined as $P_{6}=P_{4}p_{5}p_{6}=210\cdot11\cdot13=30030$.

Let us calculate the orthogonal bases of the system: $B_{1}=(1,\,0,\,0,\,0,\,0,\,0)=15015$; $B_{2}=(0,\,1,\,0,\,0,\,0,\,0)= \\ =20020$;
$B_{3}=(0,\,0,\,1,\,0,\,0,\,0)=6006$; $B_{4}=(0,\,0,\,0,\,1,\,0,\,0)=25740$; $B_{5}=(0,\,0,\,0,\,0,\,1,\,0)=16380$; $B_{6}=(0,\,0,\,0,\,0,\,0,\,1)=6930$.

Given a number $A=(1,\,2,\,2,\,3,\,6,\,4)=17$. Instead of it, after data processing we received\\ $\tilde{A}= (1,\,2,\,2,\,3,\,1,\,4)$.
To localize the error, calculate the value of the number $\tilde{A}$:
\begin{align*}
\tilde{A}=1\cdot15015+2\cdot20020+2\cdot6006+3\cdot25740+1\cdot16380+4\cdot6930-R\cdot30030=8207>210.
\end{align*}

The resulting number is incorrect ($\tilde{A}>210$), which indicates an error in the processing of data.
As a result of localization, it was determined that the number~$\tilde{\alpha}_{5}$ on the base $p_{5}=11$ was wrong.

After determining the data blocks $\tilde{M}_{i}$ and $\tilde{M}_{j}$ with broken integrity, a decision is made that
an error occurred in the sub-block $\tilde{m}_{i,\,j}$, located at the intersection of the localized row and column
of the matrix $\mathbf{\Omega^{\prime}}$ an error occurred (data integrity violation).
After localizing the error (finding the sub-block $\tilde{m}_{i,\,j}$ with integrity violation),
we perform a reconfiguration, the possibility of which is provided by redundant RNSC \cite{Yang-RNSC}.

The reconfiguration is performed by calculating $A^{*}$ from the system of equations:
\begin{align*}
    \left|A^{*}\right|_{p_{1}}=\alpha_{1},\,\cdots,\, \left|A^{*}\right|_{p_{n}}=\alpha_{n},\,\cdots,\,\left|A^{*}\right|_{p_{k}}=\alpha_{k},
\end{align*}
on the ``correct'' bases of the RNS:
\begin{align}\label{2}
    A^{*}=\left|\tilde{\alpha}_{1}B_{1,\,r}+\ldots+\tilde{\alpha}_{n}B_{n,\,r}+\ldots+\tilde{\alpha}_{k}B_{k,\,r}\right|_{P_{r}},
\end{align}
where $\tilde{\alpha}_{i}$~--- residue with error; $B_{i,\,r}$~--- orthogonal bases; $i,\,r=1,\,\ldots,\,n,\,\ldots,\,k$; $i\neq r$;
$B_{i,\,r}=\dfrac{P_{r}\mu_{i,\,r}}{p_{i}}$; $P_{r}=\dfrac{P_{k}}{p_{r}}$; $\mu_{i,\,r}$ is chosen so that the following comparison takes place: $\left|\dfrac{P_{r}\mu_{i,\,r}}{p_{i}}\right|_{p_{i}}=1$.

Let's compile Table~1 containing the values of the recalculated orthogonal bases and modules of the system,
provided that a single error occurs on each basis of the RNS, respectively.
\begin{table}[h]
\centering\label{tab:_1}
\caption{Table of values of orthogonal bases and modules of the system}
\vspace{+12pt}
  \begin{tabular}{c|cccccc}
        ~$i$~ & ~~~~$B_{1,\,r}$~~~~ & ~~~~$\cdots$~~~~ & ~~~~$B_{n,\,r}$~~~~ & ~~~~$\cdots$~~~~ & ~~~~$B_{k,\,r}$~~~~ & ~~~~$P_{r}$~~~~ \\
       \noalign{\smallskip}
        \hline
        \noalign{\smallskip}
        $1$ & $0$ & $\cdots$ & $\dfrac{P^{ }_{1}\mu_{n,\,1}}{p_{n}}$ & $\cdots$ & $\dfrac{P_{1}\mu_{k,\,1}}{p_{k}}$ & $p_{2}\ldots p_{n}\ldots p_{k}$\\ [0.25cm]
        \noalign{\smallskip}
        $\vdots$ & $\vdots$ & $\ddots$ & $\vdots$ & $\ddots$ & $\vdots$ & $\vdots$\\
        \noalign{\smallskip}
        $n$ & $\dfrac{P_{n}\mu_{1,\,n}}{p_{1}}$ & $\cdots$ & 0 & $\cdots$ & $\dfrac{P_{n}\mu_{k,\,n}}{p_{k}}$ & $p_{1}\ldots p_{n-1}p_{n+1}\ldots p_{k}$ \\ \noalign{\smallskip}
        $\vdots$ & $\vdots$ & $\ddots$ & $\vdots$ & $\ddots$ & $\vdots$ & $\vdots$\\
        \noalign{\smallskip}
        $k$ & $\dfrac{P_{k}\mu_{1,\,k}}{p_{1}}$ & $\cdots$ & $\dfrac{P_{k}\mu_{n,\,k}}{p_{n}}$ & $\cdots$ & 0 & $p_{1}\ldots p_{n}\ldots p_{k-1}$ \\ [0.25cm]
  \end{tabular}
\end{table}

After calculating $A^{*}$ on the correct bases of the system, we calculate $\alpha_{i}$ instead of the previously
excluded from the calculation residue with error $\tilde{\alpha}_{i}$:
\begin{align}\label{3}
    \alpha_{i}=\left|A^{*}\right|_{p_{i}}.
\end{align}

\textbf{Example 2}

In accordance with (\ref{2}) we calculate $A^{*}$ (\emph{the initial data from Example 1}), using Table~1, we obtain
\begin{align*}
  A^{*}=|\alpha_{1}B_{1,\,r}+\ldots+\tilde{\alpha}_{5}B_{5,\,r}+\alpha_{6}B_{6,\,r}|_{P_{5}}=|1\cdot B_{1,\,r}+\ldots+0\cdot B_{5,\,r}+4\cdot B_{6,\,r}|_{P_{5}}=17.
\end{align*}

In accordance with (\ref{3}), we calculate $\alpha_{i}$, we obtain
\begin{align*}
    \alpha_{i}=|A^{*}|_{p_{i}}=|17|_{11}=6.
\end{align*}

In the proposed system, a set of sub-blocks $m_{1,\,j},\,m_{2,\,j},\,\ldots,\,m_{n,\,j},\,m_{n+1,\,j},\,\ldots,\,m_{k,\,j}$,
which is interpreted as redundant RNSC that allow to detect an error at any stage of their processing
(provided that the multiplicity of the guaranteed error to be detected $t_{\text{det}}=d_{\min}-1$,
where $d_{\min}$~--- is the minimum code distance).

Restoration of data blocks $M_{j}^{\prime}$ in case of their integrity violation is possible by excluding
from the recovery process any $r$ sub-blocks without sacrificing the unambiguous representation
(where $r=k-n$~--- is the number of additional sub-blocks), so that the system of sub-blocks of
data blocks $M_{j}^{\prime}$ will be interpreted as nonsystematic code, or an inseparable code, and then the
sub-block is calculated $m_{i,\,j}$ instead of the previously excluded sub-block $\tilde{m}_{i,\,j}$
with the detected error.

\begin{figure}[ht]
\begin{center}
\includegraphics[height=6.8cm]{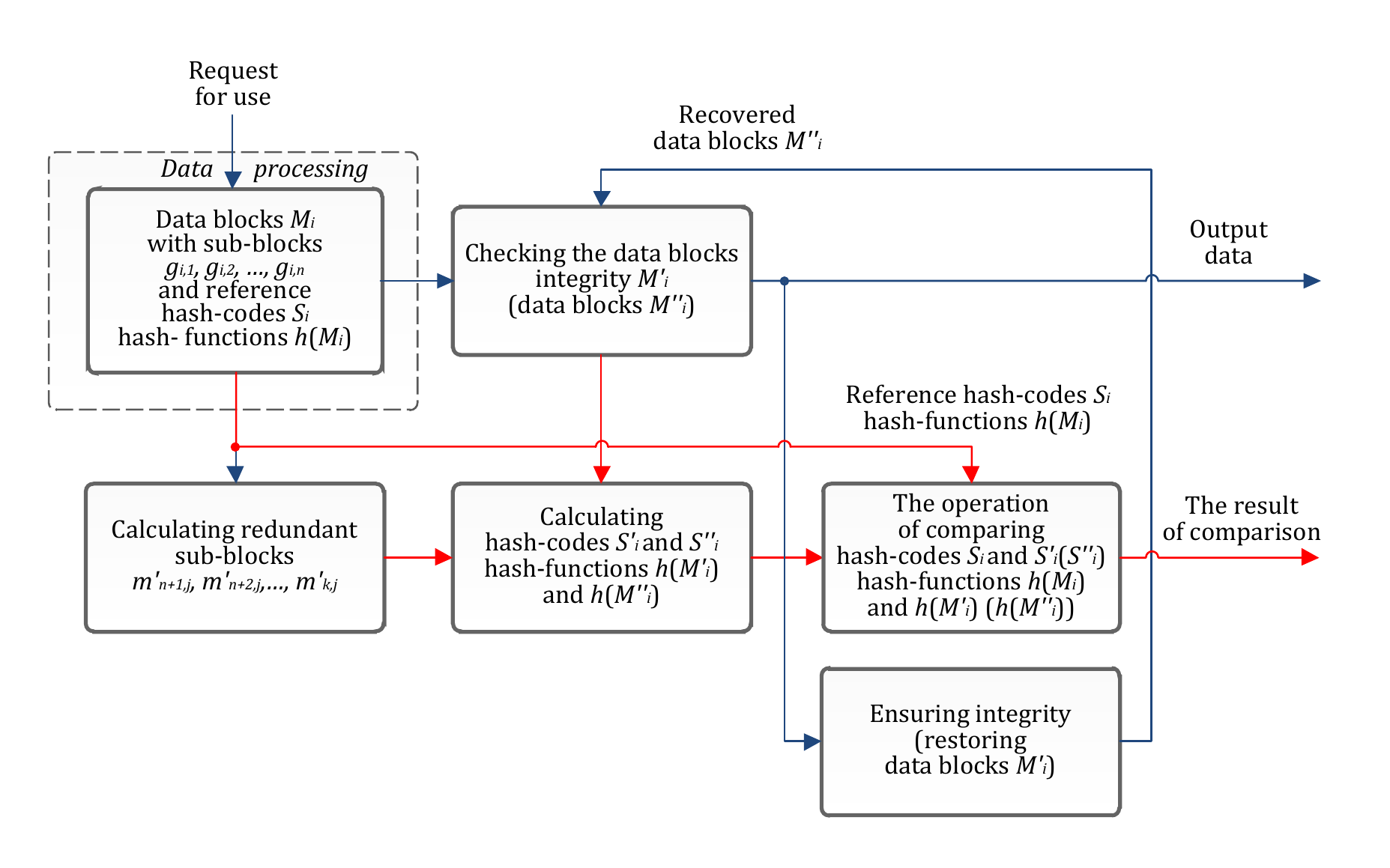}
\caption{Scheme explaining the main stage of the system construction}
\label{fig_2}
\end{center}
\end{figure}

Thus, the integrity of the data block $M_{i}$ was ensured by control and restoring the data
sub-block $\tilde{m}_{i,\,j}$ with broken integrity. Performing the verification of the data validity
(reliability, accuracy) after recovery while ensuring their integrity in case of violation
is performed by comparing the value of the previously calculated reference hash-code~$S_{i}$
hash function~$h(M_{i})$ from the data block~$M_{i}$ with the value of the calculated
hash-code~$S_{i}^{\prime\prime}$ hash function $h(M_{i}^{\prime\prime})$ already from
the restored data block~$M_{i}^{\prime\prime}$ (Figure \ref{fig_2}).

\begin{figure}[ht]
\begin{center}
\includegraphics[height=3.4cm]{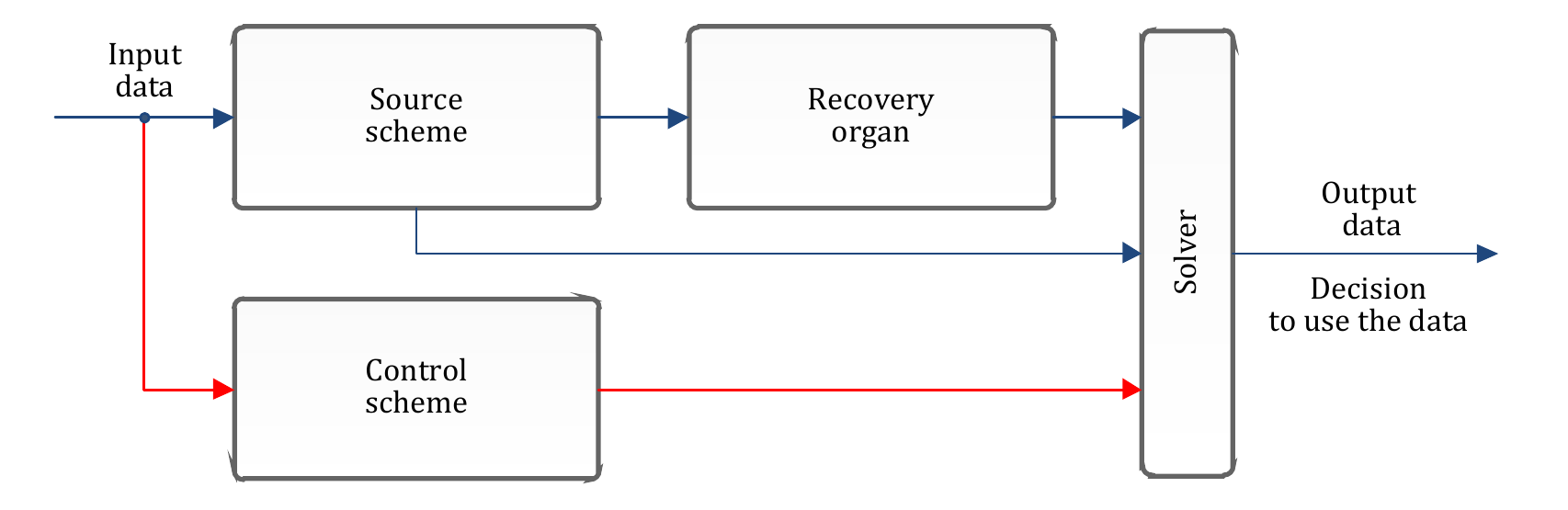}
\caption{The general scheme of the developed method}
\label{fig_3}
\end{center}
\end{figure}

The general scheme of the developed method of two-dimensional control and data integrity in information systems based on
RNSC and cryptographic hash functions is shown in the Figure \ref{fig_3}.

\section{Evaluation of the developed method}

Evaluation of the developed method is carried out in comparison with the most popular of existing solutions
integrated protection of data integrity, which consistently applies first cryptographic
data transformation to control their integrity, and then a backup technology
copy data to restore them in case of violation of integrity.

The indicator of quality is the redundancy factor $K_{\text{red}}$, which is calculated by the formula:
\begin{align}\label{4}
    K_{\text{red}}=\dfrac{V^{(\text{con})}_{\text{red.d}}+V^{(\text{ass})}_{\text{red.d}}}{V_{\text{prot.d}}},
\end{align}
where $V^{(\text{con})}_{\text{red.d}}$~--- is the amount of redundant data entered to control the integrity of the protected data,
$V^{(\text{ass})}_{\text{red.d}}$~--- is the amount of redundant data entered to assurance the integrity of the protected data,
$V_{\text{prot.d}}$~--- is the amount of data to be protected. The criterion of quality is $K_{\text{red}}\rightarrow \min$.

Since the amount of redundant data $V^{(\text{con})}_{\text{red.d}}$ introduced to control integrity in the developed method
and the existing solution are equal, then (\ref{4}) takes the form:
\begin{align}\label{5}
    K_{\text{red}}=\dfrac{V^{(\text{ass})}_{\text{red.d}}}{V_{\text{prot.d}}}.
\end{align}

In accordance with (\ref{5}) for the existing solution $K_{\text{red}}=1$,
since the amount of input redundancy is equal to the amount of data being protected ($V^{(\text{ass})}_{\text{red.d}}=V_{\text{prot.d}}$).

At the same time, to provide a level of data security,
implemented in the technology of backup, in case of violation of integrity
up to 2 sub-blocks of the data block you need to use the RNSC
with two excess bases, in this case the redundancy of the control information
is reduced from $\text{100\%}$ (with backup technology) to $\text{30-40\%}$ (RNS).

\section{Conclusion}

The results obtained provide scientific and engineering tools for control and ensuring
the data integrity with the ability to verify their validity (reliability, accuracy) after
recovery in case of violation of their integrity and provide the necessary conditions for
creating promising and improving existing information systems for various purposes.


\end{document}